\documentstyle[preprint,aps]{revtex}
\tightenlines

\begin{document}
\draft
\title{Whether can the decision of the orthopositronium problem to stimulate
studying the problem of a dark matter in the Universe?}
\author{B.M.Levin $^1$, V.I.Sokolov $^2$}
\address{A.F. Ioffe Physical-Technical Institute, 194021 St. Petersburg, Russia}
\maketitle

\begin{abstract}
The success of phenomenological model of the lifetime anomalies of the orthopositronium,
formed in substance by $\beta ^{+}$-decay positrons ($^{22}$%
Na, $^{68}$Ga, etc.), allows to assume, that the dark matter in the Universe
can be caused by realization in the $\beta ^{+}$-decay final state and in a
gravitational field of sufficient force two-digit (plus/minus) Planck mass:
vacuumlike state of matter of positive mass (on Gliner, 1965) in
limited macroscopic ``volume'' of space-time (``atom'', ``microstructure'' of vacuum:
~$\sim $ 1 km%
$^3$ during ~ $\sim \,2\cdot 10^{-6}$s) and a compensating $C$-field of
negative mass (on Hoyle \& Narlikar, 1964). For verification this hypothesis
it is necessary: for the orthopositronium problem -- to realize suggested
earlier (2005) concept of the experimentum crucis; in cosmology -- to estimate
prevalence in the Universe -- on time and place -- the $\beta ^{+}$-decay
transitions of the same type.
\end{abstract}


In this article the question on possible connection of the nature and
``microstructure'' of a dark matter in the Universe [1] with the fundamental
orthopositronium ({\it o-Ps}, $^{\text{T}}$Ps) problem [2-4] is considered.

Last work of the Michigan University group (R.S.Vallery, P.W.Zitzewitz, and
D.W.Gidley, 2003 [4]), in opinion of authors, has resolved the {\it %
o-Ps} problem. Nevertheless such decision cannot be recognized unequivocal, because does
not take into account ``isotopic anomaly'' of the {\it o-Ps} in gaseous neon [3,5].
All experimental works [4] have brought in the contribution to the
formulation of the alternative considered by us. But last result of the Michigan
group is received in qualitatively other experiment and the new experimental
technique with introduction in a measuring cell of an electric field
up to $4\cdot 10^3$V/cm has allowed to finish phenomenological
model of the {\it o-Ps} anomalies [5]. With the purpose to expand interest to the
{\it o-Ps} problem and to execute the
program of decisive experiments additional to the works [2,3,5] we consider here
possibility of its cosmological aspect.

As we shall see, the key task {\it a
priori} is an estimation of prevalence in Universe -- in a place and time --
the $\beta ^{+}$-decay transitions of the certain type.

What became a basis of the assumption of the nature of a dark matter as
``microstructures'' of vacuum, and what its connection with the {\it o-Ps}
problem?

Cosmic energy is entering into four basic components of space environment
[1]:

$\bullet $ {\it vacuum,} with relative density
$$
\Omega _{\text{V}}=\frac{\rho _{\text{V}}}{\rho _{\text{c}}}=0.7\pm 0.1;%
\eqno                       (1)
$$

$\bullet $ {\it dark substance}
$$
\Omega _{\text{D}}=\frac{\rho _{\text{D}}}{\rho _{\text{c}}}=0.3\pm 0.1;%
\eqno (2)
$$

$\bullet $ {\it shining substance of stars and galaxies} (a baryon
component)
$$
\Omega _{\text{B}}=\frac{\rho _{\text{B}}}{\rho _{\text{c}}}=0.02\pm 0.01;%
\eqno (3)
$$

$\bullet \,\,${\it ultrarelativistic environment} (radiations: photons,
neutrino)
$$
\Omega _{\text{R}}=\frac{\rho _{\text{R}}}{\rho _{\text{c}}}=0.8\cdot
10^{-5}\alpha \,\,(1<\alpha <10\div 30),\eqno (4)
$$

\noindent where $\rho _{\text{c}}$-- {\it critical density }%
$$
\rho _{\text{c}}=3H^2/8\pi G=\,(0.6\pm 0,1)\cdot 10^{-29}\text{g}\cdot \text{%
cm}^{-3}\eqno (5)
$$

\noindent and {\it H }$=65\pm 15\,$km$\cdot $s$^{\text{-1}}$Mpk$^{-1}$-- the
Hubble$^{\prime }$s constant, {\it G -- }gravitational constant.

The consideration of the dark matter nature suggested below is completely
coordinated with understanding, that the dark matter is traced by baryon
matter.

The successful phenomenology produced for a quantitative explanation of the
{\it o-Ps} anomalies [2-5], proves participation in its annihilation in the
final state of the $\beta ^{+}$-decay -- if the $\beta ^{+}$-decay occurs in a
gravitational field of sufficient force -- the {\it spacelike structures}
occupying limited macroscopic ``volume'' of space-time (~$\sim $ 1 km$^3$
during ~$\sim $ 10$^{-6}$ s), in which it is uniformity distributed
two-digit (and double!) Planck mass -- positive (``{\it microstructure}'' of
{\it vacuumlike state of matter} / VSM [6]) -- and compensating it the negative one
(the {\it C}-field [7]). In other words, with each act of the $\beta ^{+}$-decay
of the certain type nucleus in the Universe it is associated huge mass
(on the scale atomic/nuclear processes). In it will consist the basic idea of
possible participation the $\beta ^{+}$-decay nucleus of the certain type, which
involvement into formation of the dark matter of the Universe should be
estimated.

But before, we shall briefly state formulation of the question,
which is directed on performance of a pilot project on the problem {\it o-Ps}.

Phenomenological model of the {\it o-Ps } anomalies [2,3,4], based on
experiments, and formulation of a question on additional measurements [5]
has interdisciplinary physical meaning, because frameworks of consideration
of the {\it o-Ps} anomalies moves apart: instead of QED to supersymmetric
QED (SQED). This phenomenology determines the conditions of the coexistence of a
{\it close-action} (the Standard Model/SM, the ``$cG\hbar $-theory'' on
M.P.Bronshtein [8]) and non-stationary (!) {\it the long-range actions physics}
(non-Newtonian / non-Coulombian): the macroscopic spacelike vacuum structures
(discrete / crystal-like) is participate in formation of the{\it \ }%
annihilation mechanism {\it o-Ps } in different extent -- in`the resonance
conditions'' [2,3] and in not resonant conditions [4,9]). The final state of the $%
\beta ^{+}$-decay is described as the ``defect'' of the General Relativity (GR)
space-time (the limited macroscopic ``volume'') as a result of the {\it topological
quantum transition} (TQT). From the experiments position [2,3,4] and from the
phenomenology [5] our consideration is limited by TQT such as $\triangle $J$%
^\pi $ =1$^{+}$: $^{22}$Na, $^{68}$Ga, etc. The quantitative description of
{\it o-Ps } anomalies is received (the ``isotope anomaly'' [3] and
the ``$\lambda _{\text{T}}$-anomaly'' [4]) and necessity of
continuation of search of reliable criteria of the other macroscopic effects,
received in conditions having the TQT quality, -- effects is proved, which are
excluded by examination from the SM positions (in a broad sense -- from the
``$cG\hbar $- theory'' positions) [10].

In a way of receiption of positrons $p$ $\rightarrow $ $n$ $+$ $e^{+}$ $+$ $%
\nu \,\,$(in a nucleus), which then at interaction with atoms/molecules of
substance form the positronium, the nature of neutrino is attract special attention.
Still in first half 1980th, to the fiftieth anniversary of a
hypothesis of W.Pauli about neutrino, B.M.Pontekorvo did mark ``...huge
growth of the neutrino physics, which became a quantitative science, healthy and
powerful, and nevertheless promising qualitative unexpectedness'' [11].

So, the designated prospect of becoming of the ``additional $G\hbar /c$%
-physics'' is connected, first of all, to not trivial aspect of the $\beta ^{+}$%
-decay, what for the first time was showed at studying the lifetime spectra of
positrons annihilation ($^{22}$Na) in gaseous neon of the various isotope
compositions [2,3,12].

It is interesting that works, in which the mathematical bases of the
non-standard theory neutrino are submitted [13], have been published within
opening supersymmetry.

This is the abstract the first of these works:

``Among the unitary irreducible representations of the Poincare group
classified by Wigner, there occurs a class of mass-zero representations
O$_{+}^{\,\prime}(\rho )\,\,$  which has not hitherto been used for the
description of known elementary particles. For historical reasons, these are
called ``continuous-spin'' representations. They are labeled by values of
the Poincare invariants $P^2$ = 0, $W^\alpha $ $W_\alpha $= $-\rho ^2$, with
$\rho \,\,$%
\mbox{$>$}
0. {\it We propose to assign one or both neutrinos }$\nu _e${\it , }$\nu
_\mu \,\,${\it to such a representation. A corresponding neutrino can exist
in a denumerable set of helicity states , with assuming all half-odd-integer
values} $\lambda = \pm 1/2, \pm 3/2$,... . We present a theory of weak
interactions with such continuous-spin neutrinos. It differs from the
conventional $V$ $-$ $A$ theory with two-component neutrinos in the form of
the leptonic current. The predictions of the latter are approached as a
limit when $\rho \rightarrow $ 0. Explicit expressions for the matrix
elements of the new leptonic current are derived. Lepton numbers are
conserved if all the neutrinos and antineutrinos are distinct particles. It
is, however, also possible in this theory, and consistent with present
experimental evidence, to identify either $\nu _e\equiv \overline{\nu }_e$
or $\nu _\mu \equiv \overline{\nu }_\mu $. Lepton-number-nonconserving
processes are then allowed, through suppressed if $\rho \,\,$is small''.

In the more precise variant of this work ([13], 1971) the place of the
summary italicized by us is changed and added. In this last variant of the
work, unitary representations considered do not connected exclusively with
neutrino. The prospective hypothesis of their identification in the nature
is less specific and is expanded also on particles with integer helicity:

``{\it A corresponding particle can exist in a denumerable set of helicity
states} $\lambda = 0, \pm 1, \pm 2$,... {\it or all half-odd integer values}
$\lambda  = \pm 1/2, \pm 3/2$,...''. This addition has basic meaning for the
identification of the theory [13] with the orthopositronium anomalies [2-4],
undertaken by us here.

For these works [13] {\it one and only} commentary on possible connection of
these {\it lightlike} representations of the Poincare group with an
indefinite-component non-local field [14] has followed. On the other hand,
non-locality in the quantum field theory is described by {\it spacelike}
irreducible unitary representations:

``As to hyperboloids {\it p}$^2$ = C
\mbox{$<$}
0 to them too there correspond irreducible representations of the Poincare
group, but not having physical sense: on such subspace the operator of a
square of mass accepts negative value C, meanwhile as the mass of a particle
on existing representations should be non-negative'' ([15], p.205).

``These irreducible unitary representations have been associated with
``spacelike particles'' -- the so-called {\it tachyons} -- which, if they
exist, are interpreted as faster-than-light particles [16]'' [17].

The author of concept VSM [6] at the moment of its publication adhered to
such comprehension, referring to the monograph [18]: in the beginning of
chapter VII, \S 1 is marked that ``the idea of a space-time of constant
curvature
\mbox{$<$}
K%
\mbox{$>$}
of either sign is stimulating and worth exploring''. In end of \S 1 is
conclusion: ``For K
\mbox{$<$}
0, the timelike geodesics
\mbox{$<$}
...%
\mbox{$>$}
are closed curves
\mbox{$<$}
...%
\mbox{$>$}
. This depicts what can only be described as fantastic situation. We see a
test particle repeating its history over and over again! This is at variance
with our basic ideas of causality, and we conclude that a de Sitter universe
with K negative involves ideas of altogether too revolutionary a character
for physics as it exists today'' (1960!).

More profound analysis is shown that ``...there are closed timelike lines in
this space; however it is not simple connected, and if one unwraps the
circle S$^{\prime }$ (to obtain its covering space R$^{\prime }$) one
obtains the universal covering space of anti-de Sitter space which does not
contain any closed timelike lines'' [19, p.131].

The interdisciplinary physical status of the {\it o-Ps} problem is obvious,
as the phenomenology of ``additional physics'' is formulated at a level of
the groups theory, the physical theory of dimensions and structures of
superconstants of physics (${\it G}$ -- a gravitational constant, ${\it c}$ --
speed of light, $\hbar $-- Planck's constant) and of the fundamental
constants:

$\bullet \,\,\,$the fine structure constant%
$$
\alpha =\frac{e^2}{\hbar c}\,\,;
$$

$\bullet \,\,\,$the dimensionless constant of the gravitational interaction%
$$
\alpha _{\text{g}}=G\frac{m_{1\cdot }m_2}{\hbar c}\,\,;
$$

$\bullet \,\,$the ratio of proton and electron mass%
$$
\frac{m_p}{m_e}
$$

\noindent and also fundamental (plus/minus) masses: Planck mass (1899)%
$$
\text{M}_{Pl}=\pm \,\,(\hbar \cdot c)^{1/2}\cdot G^{-1/2}
$$
\noindent and Stoney mass (1881)%
$$
\text{m}_{S}=\pm \,\,e\cdot G^{-1/2}\,\,.
$$

The structure of these dimensionless constants is those that it supposes
physical interpretation both positive values of fundamental mass and their
negative values [7,20], and also negative values of two out of three
superconstants of physics -- $\hbar $[20] and $c$ [21].

It is possible to think, that similar development of physics was expected by
L.D. Landau who did not presuppose epoch of the {\it quantum chromodynamics}
(QCD). In last publication -- ``{\it About fundamental problems}'' -- Landau
has assumed ``...that the Hamilton method for strong interactions has become
obsolete and should be buried, certainly, with all honors which it has
deserved'' [22]. Becoming and development QCD was, as though, strong
counterargument. There is much speculation that in it a mistake outstanding
physics. We see in it depth of a prediction -- in fact a splintered electric
charge of quarks and confinement them are inconceivable in real life of the
observer; it is much more natural to present magnetic monopole, which will
serve a substantiation of the fundamental fact of quantization (unity) an
electric charge of all free particles [10]. If the understanding of ``{\it %
mechanisms}'' of {\it consciousness} -- one of the ``great'' problems of
physics [23] -- will demand expansion of fundamental bases, nevertheless it
will take place not in terms QCD (``color'', ``confinement'', ``asymptotic
freedom'').

Let's repeat: at any development of physics the Hamilton method
remain a basis of calculation of any ``local'' processes, and all polemic,
which has arisen on this ground can mean that the physics has achieved a
{\it point} (moment) {\it of a bifurcation,} when on base of the ``$cG\hbar $%
-theory'' and of the newest investigations and generalizations the new
physics (the ``additional $G\hbar /c$-physics'') is ``germinate''. The Hamilton
method is and remain always a basis for calculation of contributions of
elementary processes (``local'' -- in the ``$cG\hbar $-theory'') and on it
its role comes to an end, making way to an algorithm of a macroscopic ``factor
of strengthening'' [2,5,10].

The presented reasons the concept of the $C$-field with a negative energy
density [7] justify as compensating for field VSM [6], in view of what
probably ``...the simultaneous creation of quanta of fields of positive
energy field and of negative energy $C$-fields'' ([19], p.90).

In the quantum field theory the substantiation of such ``creation'' VSM ``out of
nothing'' can be achieved by development of the {\it concept of a full
relativity}: expansion of symmetry of the equations ``...up to a full
relativity, i.e. equivalence of all velocities (except the velocity of light)''
predicts existence ``...not electromagnetic long-range interactions of a
bodies with not disappearing average of a spin density. In usual conditions
to provide sublight character of relative velocities, additional symmetry
should be broken spontaneously. Restoration of
\mbox{$<$}
symmetry%
\mbox{$>$}
should be accompanied by doubling of the space-time dimension'' [24].

The full relativity in other terms ({\it a method of an} A.L.Zelmanov's {\it %
chronometric invariants}) is submitted in the theory of {\it zero-space}
(and {\it zero-particles in generalized space-time}), developed
independently as expansion of the GR space-time [25].

Suggested in [13] non-standard concept of the nature neutrino till now has
not received any identification in experiment. Publication of the $CP$%
-invariant version of the mirror universe [26], supervision of the {\it o-Ps} anomalies
in the experiments published after that [3,4] and
phenomenological model, in which the {\it o-Ps} anomaly have received a
quantitative substantiation [2,5] -- all this has created preconditions of
such identification: standard representation of the electronic neutrino is realized
during the electronic capture in atom ($K$-capture) and, probably, in the $\beta
^{+} $-decay such as $\triangle $J$^\pi $ $\neq $ 1$^{+}$, but in the $\beta ^{+}
$-decay of the nucleus such as $\triangle $J$^\pi $= 1$^{+}$ the non-standard
concept ``neutrino'' [13] are realized. In these transitions, alongside with
neutrino it is realized ``atom'' VSM with Planck mass in the limited macroscopic
``volume'' of space-time%
$$
\text{M}_{Pl}=(\hbar \cdot c)^{1/2}\cdot G^{-1/2}
$$
\noindent and the $C$-field compensating it -- the ``mirror Universe'' with
negative mass [1,4]%
$$
\text{M}_{Pl}=-\,\,(\hbar \cdot c)^{1/2}\cdot G^{-1/2}\,\,.
$$

In Standard Model the concept of the mirror world (universe) has appeared as
consequence of breaking of the $P$- and $CP$- invariance in the weak
interactions. It is appropriate to emphasize at last, that inverted commas
in our version of the ``mirror Universe'' [2,5,10,12] emphasizes indirect
cosmological aspect (mirror galaxies, stars, planets can not exist in the
nature). But can be real an ``additional cosmology'' (``microstructure'' of
vacuum).

This complex ``VSM + $C$-field'' (the ``atom of long-range action'') present
potentially in any ``point'' of space-time but realized in final state of the $%
\beta ^{+}$-decay of the nucleus such as $\triangle $J$^\pi $= 1$^{+}$,
represents of the vacuum ``microstructure''.

All told with take into account the ``mirror Universe'' is realized in
additional mode of the {\it o-Ps} annihilation [2-4] {\it o-Ps
\mbox{$\backslash$}
o-Ps}$^{^{\prime }}$({\it p-Ps}$^{^{\prime }}$) $\rightarrow \,\,\gamma
^{\circ \,}\backslash \,2\gamma ^{^{\prime }}$, where $\gamma ^{\circ \,}$--
the notoph: ``...{\it a massless particle with zero helicity, additional on the
properties to photon (helicity }$\pm $1{\it ). In interactions notoph, as
well as the photon, transfers spin} 1'' [27]. This work, though in it was
not obvious generalizations at a level of the Poincare group, it is possible to
consider as the forerunner of the works [13].

Representation of the $\beta ^{+}$-decay of the specified type now looks so $p$ $%
\rightarrow $ $n$ $+$ $e^{+}$ $+$ ``$\nu $''$\,\,$(in a nucleus + a
``macrovolume'' of space-time), where ``$\nu $ '' designates superposition
of the standard neutrino and a spacelike complex ``VSM+C-field'' that can be
characterized in terms of works [13] as ``Continuous-Spin'' Neutrino + ``
Massless Particles: the Continuous Spin Case''.

In many outlines the concept of the ``additional physics'' in the limited
``macrovolume'' of space-time has a solid-state aspect; first of all,
paradoxical realization of a {\it nuclear gamma resonance} testifies to it
in a gaseous (!) neon of the natural isotope composition that stimulated
supervision of the {\it o-Ps} ``isotope anomaly'' [2,3,5,10]: ``particles''
(positive mass $m_p$, $m_e$) and ``holes'' (negative mass -- in the ``mirror
Universe'' $-m_p$, $-m_e$) in an ``units'' of cells 3-dimensional space-like
(crystal-like) structures (the ``atom of long-range action'') -- the common
number
$$
\text{N}^{(3)}=\frac{2^{9/2}}{3\pi ^2\cdot \alpha ^9}=1.302\cdot 10^{19},%
\eqno (6)
$$

\noindent with the ``lattice'' constant%
$$
\Delta \sim c\cdot \triangle t_{\text{v}}=\frac 4{\alpha ^4}\left( \frac
\hbar {m_e\cdot c}\right) \cong 5.5\cdot 10^{-2}\text{cm}\eqno (7)
$$

\noindent ($\triangle t_{\text{v}}\,\,$-- time of existence the {\it o-Ps} in the form of
one virtual photon) and sum total mass $\text{M}_{Pl}=\sqrt{\frac{\hbar c}G}\,\,$%
plus  $\text{M}_{Pl}=-\sqrt{\frac{\hbar c}G}\,\,$and ``...a nucleus of atom of
long-range action'', containing $\overline{\text{n}}= 5.2780\cdot 10^4$
cells-''units'' (2r$_{\overline{\text{n}}}\sim 2.6\,\,$cm), which defines
the contribution of an ``additional physics'' in the {\it o-Ps} ``isotope anomaly'' in
neon (at the ``conditions of a resonance'', the factor $\sim $2) [2,3,5,10] and
in the {\it o-Ps} ``$\lambda _{\text{T}}$-anomaly'' in a not resonant conditions
(0.19$\pm $0.14)\% [4,9].

It appears also, that the ``atom of long-range action'' (the ``microstructure''
of vacuum) can be presented as an space-time ``exciton''.

Really, the antipodal
pair ``electron ($e$)-electronic hole ( $\overline{e}$)'' in the final state of the $%
\beta ^{+}$-decay together with the antipodal pair ``proton ($p$)-proton hole ($%
\overline{p}$ )'' ``annihilate'' along a symmetric variant -- ``from the leptons
up to leptons'':
$$
e\overline{e}\Leftrightarrow p\overline{p}\Rightarrow \{13\frac{\pi ^{+}\pi
^{-}}{\overline{\pi }^{-}\overline{\pi }^{+}}\,\,(\tau _{\pi ^{\pm }}\sim
2.6\cdot 10^{-8}\text{s})\rightarrow 13\frac{\mu ^{+}\mu ^{-}}{\overline{\mu
}^{-}\overline{\mu }^{+}}\,\,(\tau _{\mu ^{\pm }}\sim 2.2\cdot 10^{-6}\text{s%
})\rightarrow 13\frac{e^{+}e^{-}}{\overline{e}^{-}\overline{e}^{+}}%
\}\Rightarrow e\overline{e}\eqno (8)
$$

\noindent (brackets \{...\} include the compensating process in the ``the mirror
Universe'' -- in the ``denominators'').

This representation of model of fundamental space-like structure with
plus/minus Planck mass M$_\mu  = \pm M_{Pl}$ gives the self-coordinated
size of the ``atom of long-range action'' 2$c\cdot \tau _\mu =1.3$ km $\sim $~ 2$%
R_\mu $ and allows to offer expansion of the Huygens principle: the unit of
cellular structure of space on border of the ``atom of long-range action'', in
which there is a ``teleportation'' initial $e\overline{e}$-``pairs'', becomes
the center of the subsequent act of the ``teleportation'' -- so there is a casual
wandering such centers, i.e. the long-range action in the volume of the ``atom'' of
a defect of space-time is propagated as {\it diffusion waves}. Really, the
received estimation of the size of the ``atom'' can be presented as its ``step''
of the diffusion propagation $L_0$ = 2$R_\mu $%
$$
L_0=\sqrt{2D\tau _\mu }=\sqrt{2(L_0\cdot c)\tau _\mu },\eqno (9)
$$

\noindent where $D$ -- the diffusion coefficient. Both decisions of the
equation (9) -- $L_0$ = 0 and $L_0$ = 2$R_\mu $ mean that at the first
stage during time $\tau _\mu $ diffusion (wandering ``on a place'' owing
to ``self-promotion'' [28] with characteristic time 1/$\Omega \sim $ 10$%
^{-44}$s) should have character of casual rotation in the volume of the ``atom''.
Plus/minus Planck mass%
$$
\text{M}_{Pl}=\pm (\hbar c)^{1/2}\cdot G^{-1/2}\equiv 2.177\cdot 10^{-5}%
\text{g\ ,}
$$

\noindent it is submitted, according to (6), through the fine structure
constant $\alpha \,\,$and masses $m_p$ and $m_e$ [5]%
$$
\text{M}_\mu =\text{N}^{(3)}\cdot (m_p+m_e)=\frac{2^{9/2}}{3\pi ^2\alpha ^9}%
\cdot (m_p+m_e)\equiv 2.179\cdot 10^{-5}\text{g\ .}
$$
\noindent It means, that each of $\sim $10$^{19}$ cells of spacelike
structure is displayed in each of $\sim $10$^{19}$ ``mirror'' cells on the
mechanism of a tachyon self-promotion [28,2,12]. At the second stage
diffusion on distance $L$ to the moment $t$ ($t$ = 0 in moment of the $\beta
^{+} $-decay) takes place%
$$
L=2\sqrt{\frac t{\tau _\mu }}R\,\,.
$$

As it is already noticed above, such discrete structure in the limited
``volume'' of space-time (the ``atom of long-range action'') on background GR
can be submitted as the ``defect'' of space-time, which is formed in a the final
state of {\it topological quantum transition}. If the ``atom of long-range
action'' at the moment of a birth in the final state of the $\beta ^{+}$-decay or as
a result of the subsequent diffusion appears in a gravitational field with
critical value of the free falling acceleration $\gamma _{cr}$ there is its
``splitting'' on the plus/minus Planck masses (the {\it generalized displacement
current}), i.e. in one-stage (from ``nothing''!) ``elementary'' macroscopic
``domain'' of a dark matter with the mass 2$\left| \text{M}_{Pl}\right| $ is
born. The critical acceleration we shall estimate, using the Compton length of a
wave of the proton and the ``time of rest'' (the time of one ``step'') of the ``atom of
long-range action'' [5]%
$$
\gamma _{cr}=\frac \hbar {m_p\cdot c\cdot \tau _\mu ^2}=\frac{\hbar c}{%
m_p\cdot R_\mu ^2}\sim 0.01\,\,\text{cm/s}^2.\eqno (10)
$$

Let's carry out the necessary estimations and, as far as possible, we compare
them with the available cosmological data on the contribution of a dark matter.
So, from (6) and (7) it is received an absolute value of the ``baryon matters''
density (VSM + $C$-field) in the final state of the $\beta ^{+}$-decay of a nucleus
in the volume $V_\mu $ = N$^{(3)}\cdot \Delta ^3$ and in the gravitational field of
sufficient force%
$$
\rho _{\left| \pm \text{M}_{Pl}\right| }=\frac{2\text{M}_{Pl}}{V_\mu }\cong
2\cdot 10^{-20}\text{g}\cdot \text{cm}^{-3}\,\,.\eqno (11)
$$

The density (11) should be compared to the value of the density of a dark matter
accepted for today in the Universe. From (2) and (11) it is received%
$$
\rho _{\text{``D''}}=f_{\text{U}}\cdot \rho _{\left| \pm \text{M}%
_{Pl}\right| }\sim \Omega _{\text{D}}\cdot \rho _{\text{c}}\,\,,
$$

\noindent where $f_{\text{U}}$-- the ``universal'' factor, which is taking
into account the all sources and the physical (an ``additional physics'') mechanisms of
participation of the spacelike structure accompanying each act of the $\beta
^{+} $-decay in the final state in view of a possible mechanisms of its
``dissipation'' or disappearance. As shown above, the spacelike structure
in the final state of the $\beta ^{+}$-decay can diffusion from areas of the
Universe, where gravitation is enough for display $\pm $M$_{Pl}\Rightarrow
2\left| \text{M}_{Pl}\right| $, into the ``free Cosmos'', where the mass of the
spacelike structure is defined by Heisenberg uncertainty relations%
$$
m_\mu \cong \frac \hbar {2R_\mu \cdot c}\sim 10^{-10}\text{eV,}
$$

\noindent and again in those areas of Cosmos, where by criterion (10) its
mass $2\left| \text{M}_{Pl}\right| $ will be shown; the index in $\rho _{%
\text{''D''}}$ reflects hypothetical character of an estimation. We receive%
$$
f_{\text{U}}\sim \frac{\Omega _{\text{D}}\cdot \rho _{\text{c}}}{\rho
_{\left| \pm \text{M}_{Pl}\right| }}\sim 10^{-10}\,\,.\eqno (12)
$$

It is visible, as far as it is complex and, probably, even unattainable the
task of a full substantiation of an estimation $f_{\text{U}}$, but it does
not change the basic attitude to the considered mechanism of a dark matter
formation. The reality of this program can be confirmed or denied as a
result of realization {\it experimentum crucis} for the decision of the {\it %
orthopositronium problem} [5] and forthcoming {\it cosmological
(astrophysical) estimations of the ``universal'' factor} $f_{\text{U}}$.%
\newline

\noindent $^1$ E-mail address: bormikhlev@mail.ioffe.ru , bormikhlev@mail.ru \newline
$^2$ E-mail address: v.sokolov@mail.ioffe.ru

\end{document}